\documentclass[twocolumn,english,conference]{IEEEtran}
\usepackage[T1]{fontenc}
\usepackage{babel}
\usepackage{amsmath}
\usepackage{amsthm}
\usepackage{amssymb}
\usepackage{graphicx}
\usepackage[unicode=true,
 bookmarks=true,bookmarksnumbered=true,bookmarksopen=true,bookmarksopenlevel=1,
 breaklinks=false,pdfborder={0 0 0},pdfborderstyle={},backref=false,colorlinks=false]
 {hyperref}
\hypersetup{pdftitle={Your Title},
 pdfauthor={Your Name},
 pdfpagelayout=OneColumn, pdfnewwindow=true, pdfstartview=XYZ, plainpages=false}

\makeatletter
\theoremstyle{plain}
\newtheorem{thm}{\protect\theoremname}
\theoremstyle{plain}
\newtheorem{cor}[thm]{\protect\corollaryname}
\theoremstyle{plain}
\newtheorem{prop}[thm]{\protect\propositionname}
\theoremstyle{remark}
\newtheorem{rem}[thm]{\protect\remarkname}

\IEEEoverridecommandlockouts

\makeatother

\providecommand{\corollaryname}{Corollary}
\providecommand{\propositionname}{Proposition}
\providecommand{\remarkname}{Remark}
\providecommand{\theoremname}{Theorem}

\begin{document}

\title{Repair of Multiple Descriptions on Distributed Storage}

\author{\IEEEauthorblockN{Anders H{\o}st-Madsen}\IEEEauthorblockA{Department of Electrical Engineering\\
University of Hawaii, Manoa\\
Honolulu, HI, 96822, Email: ahm@hawaii.edu}\and\IEEEauthorblockN{Heechoel Yang, Minchul Kim, Jungwoo Lee}\IEEEauthorblockA{Department of Electrical and Computer Engineering\\
Seoul National University \\
E-mail: \{hee2070,kmc1222\}@cml.snu.ac.kr, junglee@snu.ac.kr}}
\maketitle
\begin{abstract}
In multiple descriptions on distributed storage, a source is stored
in a shared fashion on multiple servers. When a subset of servers
are contacted, the source should be estimated with a certain maximum
distortion depending on the number of servers. The problem considered
in this paper is how to restore the system operation when one of the
servers fail and a new server replaces it, that is, repair. The requirement
is that the distortions in the restored system should be no more than
in the original system. The question is how many extra bits are needed
for repair. We find the optimum solution for a two server problem
in the Gaussian case, and an achievable rate for general $n$ nodes.
One conclusion is that it is necessary to design the multiple description
codes with repair in mind; just using an existing multiple description
code results in unnecessary high repair rates.
\end{abstract}

\section{Introduction}

In distributed storage systems \cite{DimakisWainwrightAl10} information
is stored in a shared fashion among multiple servers; to recover the
information, in principle all servers are contacted and the information
combined. Multiple description coding \cite{ElGamalCover82} can be
seen as a variation of distributed storage. When only some servers
are contacted, instead of a failure, a distorted version of the information
is recovered. As more servers are contacted, the distortion can be
reduced. This can for example be used for distributed storage of video.

A central issue in distributed storage is how to repair the system
when one or more of the servers fail or become unavailable \cite{DimakisWainwrightAl10}
and is replaced by new servers. In traditional distributed storage
this is in principle easily solved by using minimum distance separable
(MDS) $(n,k)$ erasure codes: when $k$ out of $n$ servers are available,
the total information can be recovered, and a new $(n,k)$ erasure
code generated. With multiple description coding this is not a feasible
solution: when less than $k$ servers are available with an $(n,k)$
erasure code, no information can be recovered, even with distortion
\textendash{} as also explained in \cite{PuriRamchandranAl04PartI}. 

This leads us to the problem we consider in this paper. A multiple
description coding system is specified as follows: when a subset $S\subset\{1,\ldots,n\}$
of servers are contacted, a source $X$ should be restored with a
distortion at most $D(S)$. Now, if one (or multiple) of the servers
fail, we should be able to set up a replacement server with enough
information so that the whole region $D(S),S\subset\{1,\ldots,n\}$
is restored. There are a number of variations of this problem
\begin{enumerate}
\item There is special (highly reliable) repair server that does not participate
in the usual operation of the system, but only comes into action if
another server fails. Either
\begin{enumerate}
\item The repair server can contact all other (non-failed) servers and
use their information combined with its own information to restore
the failed server (collaborative repair).
\item The repair server has to restore the failed server without contacting
other servers (non-collaborate repair).
\end{enumerate}
\item The repair information is stored in a distributed fashion among the
$n$ servers. Now it is immediately clear that $D(\{1,\ldots,n\})$
cannot be restored. We can therefore consider two cases
\begin{enumerate}
\item No more than $n-1$ servers are ever accessed for normal operation,
and the problem is to restore $D(S)$ for $|S|\leq n-1$. 
\item We require restoring the total region $D(S)$ for all $S$. The only
solution here is to modify the system so that $D(S)=D(\{1,\ldots,n\})$
for all $S$ with $|S|=n-1$. The technical solution is therefore
exactly the same as above. The only difference is how to do the accounting.
\end{enumerate}
\end{enumerate}
The problem we set out to solve is to find the minimum information
that can be stored on the servers so that the above constraints can
be satisfied.

The problem of repair of multiple descriptions has been considered
in a number of previous papers. In \cite{ChanHo13} the authors consider
a problem like 1. above, but they do not give a single letter description
of rate regions. In \cite{KapetanovicOtterstenAl14} the authors consider
practical codes for repairing. In the current paper we aim to provide
single letter expression for achievable rate regions, and in some
cases the actual rate region.

In the following we use the term \emph{repair node} for the special
repair server and \emph{operational nodes }to denote the other servers.
We use $I_{k}=\{1,\ldots,k\}$, which used as an index means $X_{I_{k}}=[X_{1},\ldots,X_{k}]$.

\section{\label{Problem.sec}Problem Description}

We consider a symmetric multiple description problem as in \cite{PuriRamchandranAl05PartII},
and use their notation. We have an i.i.d. (independent identically
distributed) source $X$ that takes values in a finite alphabet $\mathcal{X}$
and needs to be restored in the alphabet $\hat{\mathcal{X}}$, with
generalizations to a Gaussian source through usual quantization arguments
\cite{ElGamalKimBook}. We will first define the distributed repair
problem. For a source sequence $x^{l}$ of length $l$ each node stores
$lR_{t}$ bits. For many achievable schemes, these can be split into
$lR$ bits for normal operation and $lR_{r}$ additional bits used
only for repair. There are $n$ encoding functions $f_{i}:\mathcal{X}^{l}\to\{1,\ldots,2^{lR_{t}}\}$
, $2^{n-1}$ decoding function $g_{J}:\{1,\ldots,2^{lR_{t}}\}^{|J|}\to\mathcal{\hat{X}}^{l}$,
$J\subset I_{n}$, and $n$ repair functions $h_{i}:\{1,\ldots,2^{lR_{t}}\}^{n-1}\to\{1,\ldots,2^{nR_{t}}\}$.
We define the error probability of repair as
\[
P_{r}^{(l)}=\max_{i=1,\ldots,n}P\left(h_{i}(f_{I_{n}-\{i\}}(x^{l}))\neq f_{i}(x^{l})\right)
\]
We now say that a a tuple $(R_{t},D_{1},\ldots,D_{n-1})$ is achievable
if there exists a sequence of $(2^{lR_{t}},l)$ codes with
\begin{align}
\lim_{l\to\infty}\max_{J:|J|=m}E[d_{|J|}(x^{l},g_{J}(f_{J}(x^{l})))] & \leq D_{m}\nonumber \\
\lim_{l\to\infty}P_{r}^{(l)} & =0\label{RepairPerf.eq}
\end{align}
with the distortions $d_{|J|}(x^{l},\hat{x}^{l})=\frac{1}{l}\sum_{i=1}^{l}\tilde{d}_{|J|}(x_{i},\hat{x}_{i}),\tilde{d}_{|J|}(x_{i},\hat{x}_{i})\geq0$.
We call this \emph{exact repair}. The repaired node is required to
be an exact copy of the failed node, except that we allow a certain,
vanishing, error rate. Notice that the randomness in the system is
purely due to the source $x^{l}$. Thus, for a given sequence $x^{l}$
either all failures can be repaired exactly, and if they can be repaired
once, they can be repaired infinitely many times; or, some failures
can never be repaired. The probability of the source sequences that
are not repairable should be vanishing small.

An alternative problem formulation, which we  call \emph{functional
repair}, is to allow approximate repair, where the only requirement
is that after repair the distortion constraint is satisfied. In that
case one would have to carefully consider repeated repair. In this
paper we will only consider strong repair.

For a dedicated repair node, each node stores $lR$ bits and the repair
node $lR_{r}$ bits. The non-collaborative repair functions are instead
functions $h_{i}:\{1,\ldots,2^{lR_{r}}\}\to\{1,\ldots,2^{lR}\}$,
and the collaborative repair functions are $h_{i}:\{1,\ldots,2^{lR_{r}}\}\times\{1,\ldots,2^{lR}\}^{n-1}\to\{1,\ldots,2^{lR}\}$,
with the other definitions similar.

\section{\label{TwoNodes.sec}Two Nodes }

We at first consider a problem with $n=2$ nodes as this is one of
the only cases where the optimum rate distortion region is known,
in the Gaussian case \cite{ElGamalCover82} with mean-squared distortion.
If there is no repair node, the problem is trivial: each node has
to be able to achieve the distortion $D_{2}$ by itself, and they
can therefore be copies of each other. We therefore assume that there
is a special repair node, and consider the case when this has access
to the surviving operational nodes for repair, collaborative repair.
The question is: what is the minimum information the repair node can
store, so that $(D_{1},D_{2})$ can be achieved \emph{without any
increase in storage rate of the operational nodes}. 

The problem is most well posed in the Gaussian case, as we know the
exact rate distortion region $(R,D_{1},D_{2})$. We then want to find
the minimum repair rate $R_{r}$ for every point on the boundary of
the rate distortion region $(R,D_{1},D_{2})$. We also know that the
El-Gamal Cover (EC) coding scheme achieves the optimum rate-distortion
region. The idea in the EC scheme is that each node stores an index
for use when only that node is accessed, in addition to half the bits
of an index with refinement information that is only used when both
nodes are accessed. However, EC is clearly sub-optimum for repair.
Consider the point $D_{2}=D_{1}$; in this point it is clear what
is the optimum solution. Each node has to be able to restore the source
by itself with distortion $D_{2}$, and they can therefore be copies
of each other. Repair then is done simply by copying from the surviving
node and $R_{r}=0$. Now if $D_{2}$ is close to $D_{1}$ one would
expect $R_{r}$ to be small. On the other hand, in EC the two nodes
store independently generated codewords \textendash{} even if the
joint distribution is not independent \cite{ElGamalCover82}. Therefore
to restore the EC code exactly, $R_{r}=2R$ is needed.

We therefore instead consider the Zhang-Berger (ZB) scheme \cite{ZhangBerger87,ElGamalKimBook}.
In addition to the individual and refinement information stored in
the EC scheme, the nodes in the ZB scheme store a common codeword.
While this cannot decrease rate in the Gaussian case, a common codeword
is great for repair, as it can be simply copied from the surviving
node without additional information from the repair node.

Instead of the original characterization of the ZB scheme, we will
describe it in the language of PRP \cite{PuriRamchandranAl05PartII},
both to be consistent with the general problem later, and because
the PRP more explicitly characterizes the information stored on nodes
in terms of auxiliary random variable, which is essential to calculate
repair rate.
\begin{thm}[Zhang-Berger]
A rate $R$ is achievable if
\begin{align*}
R & >I(X;U_{1})+H(Y_{12}|U_{1})+\frac{1}{2}H(Y_{2}|Y_{12},Y_{11},U_{1})\\
 & -\frac{1}{2}H(Y_{12},Y_{11}|X,U_{1})-\frac{1}{2}H(Y_{2}|Y_{12},Y_{11},X,U_{1})
\end{align*}
for some conditional pdf $p(u_{1},y_{11},y_{12},y_{2}|x)$ such that
$E[d_{1}(X,g_{1i}(U_{1},Y_{1i}))]\leq D_{1}$, $E[d_{2}(X,g_{2}(U_{1},Y_{1i}))]\leq D_{2}$. 
\end{thm}
\begin{cor}
A repair rate $R_{r}$ is achievable if
\begin{align}
R_{r} & >H(Y_{12}|Y_{11},U_{1})-\frac{1}{2}H(Y_{12},Y_{11}|X,U_{1})\nonumber \\
 & +\frac{1}{2}H(Y_{2}|Y_{12},Y_{11},U_{1})-\frac{1}{2}H(Y_{2}|Y_{12},Y_{11},X,U_{1})\label{Rr2nodeg.eq}
\end{align}
\end{cor}
We omit the proof, as it is a special case of Theorem \ref{CollaborativeRepair.thm}
later., but we will briefly outline how the repair works. First, the
common codeword needs no extra information for repair. For the base
layer, there are about $l(H(Y_{12}|U_{1})-\frac{1}{2}H(Y_{12},Y_{11}|X,U_{1}))$
bits stored on each node. Suppose it is known in advance that node
2 will fail. The repair node could simply store a copy of the $l(H(Y_{12}|U_{1})-\frac{1}{2}H(Y_{12},Y_{11}|X,U_{1}))$
bits of node 2. But at the time of repair, the codeword in node 1
is known. We can use that to reduce the number of required bits for
repair to $l(H(Y_{12}|Y_{11},U_{1})-\frac{1}{2}H(Y_{12},Y_{11}|X,U_{1}))$
(the proof will make more clear how this works). This gives the first
term in (\ref{Rr2nodeg.eq}). It is of course not known in advance
which node will fail, but this can be solved through binning (think
an $(2,1)$ erasure code) without extra rate. The resolution information
requires about $l(H(Y_{2}|Y_{12},Y_{11},U_{1})-H(Y_{2}|Y_{12},Y_{11},X,U_{1}))$
bits. Each operational node stores half the bits. The repair node
can then for example store the xor of the two sequences of bits, so
that the lost sequence can be recovered when the other sequence is
known; this gives the second term of (\ref{Rr2nodeg.eq}).

It turns out the ZB is exactly optimum in the Gaussian case
\begin{thm}
\label{2node.thm}Consider a Gaussian source with $E[X^{2}]=1$. ZB
achieves the following repair rate
\[
R_{r}=\begin{cases}
\frac{1}{4}\log\left(\frac{1}{D_{2}}\right) & D_{2}\leq2D_{1}-1\\
\frac{1}{2}\log\left(\frac{2\sqrt{(1-D_{1})(D_{1}-D_{2})}}{(D_{2}-1)\sqrt{D_{2}}}\right) & 2D_{1}-1\leq D_{2}\leq\frac{D_{1}}{2-D_{1}}\\
\frac{1}{2}\log\left(\frac{D_{1}}{D_{2}}\right) & \frac{D_{1}}{2-D_{1}}\leq D_{2}
\end{cases}
\]
This is the optimum repair rate.
\end{thm}
\begin{IEEEproof}
For achievable rate we let $U_{1}=X+Q_{u1},Y_{1i}=X+Q_{1i},Y_{2}=X+Q_{2}$
with $Q_{\ldots}$ zero-mean Gaussian, $E[Q_{u1}^{2}]=\sigma_{u1}^{2}$,
$E[Q_{1i}^{2}]=\sigma_{q1}^{2}$, $E[Q_{2}^{2}]=\sigma_{q2}^{2}$,
$E[Q_{11}Q_{12}]=\rho_{1}\sigma_{q1}^{2}$, and all other noise variables
uncorrelated. We first calculate the distortions,
\begin{align}
D_{1} & =\frac{\sigma_{q1}^{2}\sigma_{u1}^{2}}{\sigma_{q1}^{2}\sigma_{u1}^{2}+\sigma_{q1}^{2}+\sigma_{u1}^{2}}\nonumber \\
D_{2} & =\frac{(\rho_{1}+1)\sigma_{q1}^{2}\sigma_{q2}^{2}\sigma_{u1}^{2}}{(\rho_{1}+1)\sigma_{q1}^{2}\left(\sigma_{q2}^{2}\sigma_{u1}^{2}+\sigma_{q2}^{2}+\sigma_{u1}^{2}\right)+2\sigma_{q2}^{2}\sigma_{u1}^{2}}\label{D22node.eq}
\end{align}
The $D_{1}$ distortion constraint is always satisfied with equality,
and therefore
\begin{equation}
\sigma_{q1}^{2}=\frac{D_{1}\sigma_{u1}^{2}}{\sigma_{u1}^{2}-D_{1}\sigma_{u1}^{2}-D_{1}}\label{sq1s.eq}
\end{equation}

Using standard Gaussian calculations of differential entropy, we get

\begin{align}
\lefteqn{R=\frac{1}{2}\log\left(1+\frac{1}{\sigma_{u1}^{2}}\right)+\frac{1}{2}\log\left(\frac{\sigma_{u1}^{2}}{\sqrt{1-\rho_{1}^{2}}D_{1}\left(\sigma_{u1}^{2}+1\right)}\right)}\nonumber \\
 & +\frac{1}{4}\log\left(\!\frac{D_{1}\left((\rho_{1}\!-\!1)\sigma_{q2}^{2}\left(\sigma_{u1}^{2}\!+\!1\right)\!+\!(\rho_{1}\!+\!1)\sigma_{u1}^{2}\right)\!+\!2\sigma_{q2}^{2}\sigma_{u1}^{2}}{\sigma_{q2}^{2}\left(D_{1}(\rho_{1}\!-\!1)\left(\sigma_{u1}^{2}\!+\!1\right)\!+\!2\sigma_{u1}^{2}\right)}\!\right)\label{R2node.eq}
\end{align}
and 
\begin{align}
R_{r} & =\frac{1}{2}\log\left(\frac{1-\rho_{1}}{\sqrt{1-\rho_{1}^{2}}\sigma_{q2}^{2}\sigma_{u1}^{2}}\right)+\frac{1}{2}\log\left(2\sigma_{q2}^{2}\sigma_{u1}^{2}\right.\nonumber \\
 & \left.D_{1}\left((\rho_{1}-1)\sigma_{q2}^{2}\left(\sigma_{u1}^{2}+1\right)+(\rho_{1}+1)\sigma_{u1}^{2}\right)\right)\label{Rr2node.eq}
\end{align}

Following \cite[Theorem 13.2]{ElGamalKimBook}, there are three regions
for $D_{2}$ to consider. If $D_{2}\geq2D_{1}-1$, the optimum solution
can be achieved without transmitting resolution information, i.e.,
$\sigma_{q2}^{2}=\infty$. From (\ref{R2node.eqx}) we get
\begin{equation}
R=\frac{1}{2}\log\left(\frac{1}{D_{1}\sqrt{1-\rho_{1}^{2}}}\right)\label{RregionI.eq}
\end{equation}
independent of $\sigma_{u1}^{2}$. This region is again split into
two. If $D_{2}>\frac{D_{1}}{2-D_{1}}$ we can achieve $R=\frac{1}{2}\log\left(\frac{1}{D_{1}}\right)$,
which is achieved (and only achieved) for $\rho_{1}=0$. What happens
in this region is that the two nodes have independent messages, and
the combination results in a distortion less than $D_{2}$. But independent
messages are poor for repair. We cannot change $\rho_{1}$ because
of (\ref{RregionI.eq}), but we can use the common message in the
ZB scheme. We choose the power $\sigma_{u1}^{2}$ so that the combination
of the two nodes' information gives exactly a distortion $D_{2}$,
which gives $\sigma_{u1}^{2}=\frac{D_{1}D_{2}}{2D_{2}-D_{1}D_{2}-D_{1}}$.
This solution is valid for $D_{2}>\frac{D_{1}}{2-D_{1}}$. We then
get from (\ref{Rr2node.eq}) that
\[
R_{r}=\frac{1}{2}\log\left(\frac{D_{1}}{D_{2}}\right)
\]
For the case $D_{2}\leq\frac{D_{1}}{2-D_{1}}$ we need to decrease
$\rho_{1}$ from zero. We store no common message. Then, solving (\ref{D22node.eq})
with respect to $\rho_{1}$ (for $\sigma_{q2}^{2}=\infty$ and $\sigma_{u1}^{2}=\infty$
) gives $\rho_{1}=\frac{D_{1}D_{2}+D_{1}-2D_{2}}{D_{1}(D_{2}-1)}$
and
\begin{align*}
R & =\frac{1}{2}\log\left(\frac{D_{2}-1}{2\sqrt{(D_{1}-1)D_{2}(D_{2}-D_{1})}}\right)\\
R_{r} & =\frac{1}{2}\log\left(\frac{2\sqrt{(D_{1}-1)(D_{2}-D_{1})}}{(D_{2}-1)\sqrt{D_{2}}}\right)
\end{align*}

In the region $D_{2}\leq2D_{1}-1$ the optimum solution requires storage
of resolution information; we use no common message. We get 
\[
R=\frac{1}{2}\log\left(\sqrt{\frac{1}{D_{1}D_{2}(1-\rho_{1})(D_{1}(\rho_{1}-1)+2)}}\right)
\]
We minimize this with respect to $\rho_{1}$ and get $\rho_{1}=\frac{D_{1}-1}{D_{1}}$.
Inserting this we get
\begin{align*}
R & =R_{r}=\frac{1}{2}\log\left(\frac{1}{\sqrt{D_{2}}}\right)
\end{align*}

For the converse, we can think of the problem as follows. When the
repair node has restored the failing operational node, the two operational
nodes should be able to estimate $X$ with a distortion\footnote{Notice that this is a much weaker requirement than (\ref{RepairPerf.eq})
and corresponds more or less to functional repair in Section \ref{Problem.sec}.} (less than or equal to) $D_{2}$. But that also means that the surviving
node and the repair node when they cooperate must be able to estimate
$X$ with a distortion $D_{2}$. From standard rate-distortion theory
we then must have $R+R_{r}\geq\frac{1}{2}\log\left(\frac{1}{D_{2}}\right)$.
Now it is easy to see that in all three regions above, we have $R+R_{r}=\frac{1}{2}\log\left(\frac{1}{D_{2}}\right)$.
\end{IEEEproof}

\section{General $n$ nodes}

For more than two nodes the optimum rate distortion region is not
known, not even in the Gaussian case. There are therefore many different
schemes for multiple description coding, e.g., \cite{VenkataramaniKramerAl03,TianChen10,ViswanathaRoseAl16,PuriRamchandranAl05PartII},
and we have to design repair for each specific method. In this paper
we will consider the PRP scheme \cite{PuriRamchandranAl04PartI,PuriRamchandranAl05PartII},
as this is specifically aimed at the symmetric case and is well-suited
to repair. While there are certain cases where the PRP scheme can
be improved \cite{TianChen10}, these schemes are much more involved
and do not universally improve rate.

Let us briefly outline the standard PRP scheme. In layer $k$, $n$
codewords $Y_{kI_{n}}$

To simplify the discussion, consider $n=3$ nodes. The problem is
specified by the distortions $(D_{1},D_{2})$. As in the two node
case, there is one point where we know the optimum solution: if $D_{2}=D_{1}$
the optimum solution is to let all three nodes be identical copies,
so that $R=\frac{1}{2}\log\left(D_{1}^{-1}\right)=\frac{1}{2}\log\left(D_{2}^{-1}\right)$.
The solution is continuous in this point: if $D_{2}=D_{1}-\epsilon$,
we can store identical codewords in the tree nodes, in this case so
that each can individually restore to $D_{2}$, with a rate $R=\frac{1}{2}\log\left(D_{2}^{-1}\right)$.
There is no claim that this is optimum, but it is better than the
PRP solution: For $D_{2}=D_{1}-\epsilon$ the PRP solution is to store
completely independent codewords in the three nodes ($p(y_{11},y_{12},y_{13}|x)=p(y_{11}|x)p(y_{12}|x)p(y_{13}|x)$);
when they are combined the achieve less than $D_{2}$ distortion.
Independent codewords are poor for repair, since independent redundant
copies must be stored. One can improve repairability by choosing correlated
codewords $Y_{11},Y_{12},Y_{13}$, but not much. The issue is that
the codebooks $\mathcal{C}_{1i}$ of size $2^{lR'}$ for $Y_{1i}$
in \cite[Section III.D]{PuriRamchandranAl05PartII} are generated
independently according to the marginal distribution $p(y_{11})$.
At encoding, the encoder finds $n$ codewords that are \emph{jointly}
typical with $x^{l}$. If the joint distribution is highly dependent,
there are not many such codewords: according to \cite[(11)]{PuriRamchandranAl04PartI}
we must have $nR'>\sum_{i=1}^{n}H(Y_{i})-H(Y_{1},\ldots,Y_{n}|X)$;
if the $Y_{i}$ are highly dependent (good for repair), the difference
is large. This is not specific to PRP, it is a common feature of all
multiple description coding schemes.

A solution is to allow common messages, as in ZB and other schemes
\cite{VenkataramaniKramerAl03,ViswanathaRoseAl16}. This can seems
like a crude solution, but we know that this was exactly optimum in
the two node case. We are not claiming that this will improve PRP
as such, although it could, but it will improve repairability.

As baseline, consider the standard PRP scheme where we use at most
$n-1$ nodes for the reconstruction. Now in layer $n-1$, we just
need a single common message (in standard PRP that happens at layer
$n$). This message can be encoded using an $(n,n-1)$ MDS erasure
code. We then get the following rate which we state without proof
as it is a simple modification of PRP
\begin{prop}
\label{PRPmodfied.thm}The following rate is achievable with $n$
nodes and using at most $(n-1)$ nodes for reconstruction
\begin{align*}
R & \geq H(Y_{1n})+\sum_{k=2}^{n-2}\frac{1}{k}H(\mathbf{Y}_{kI_{k}}|\mathbf{Y}_{I_{k-1},I_{k}})\\
 & +\frac{1}{n-1}H(Y_{n-1}|\mathbf{Y}_{I_{n-2}I_{n-1}})-\frac{1}{n}H(\mathbf{Y}_{I_{n-2}I_{n}}|X)\\
 & -\frac{1}{n-1}H(Y_{n-1}|\mathbf{Y}_{I_{n-2}I_{n-1}},X)
\end{align*}
\end{prop}
Repair is done layer-by-layer. In each layer, in addition to the standard
PRP codewords, we allow for a common codeword shared among all nodes,
and encoded with an $(n,k)$ erasure code, since a common codeword
is good for repair, as seen in the two node case. We now have the
main result

\begin{thm}[Distributed repair]
\label{DistributedRecov.thm}For any symmetric probability distribution
\cite{PuriRamchandranAl05PartII} $p(\mathbf{y}_{I_{n-2},I_{n}},\mathbf{u}_{I_{n-2}},y_{n-1}|x)$
and decoding functions $g_{J}$ the lower convex closure of $(R+R_{r},D_{1},\ldots,D_{n-1})$
is achievable, where $E[d_{|J|}(X,g_{J}(\mathbf{Y}_{I_{|J|}J},\mathbf{U}_{I_{|J|}})]\leq D_{|J|},|J|\leq n-1$
and the information needed to encode operational information is
\begin{align*}
\lefteqn{R>I(X;U_{1})+H(Y_{1n}|U_{1})+\sum_{k=2}^{n-2}\frac{1}{k}H(\mathbf{Y}_{kI_{k}}|\mathbf{Y}_{I_{k-1},I_{k}}\mathbf{U}_{I_{k}})}\\
 & +\frac{1}{n-1}I(Y_{n-1};X|\mathbf{Y}_{I_{n-2}I_{n-1}},\mathbf{U}_{I_{n-2}})\\
 & -\frac{1}{n}H(\mathbf{Y}_{I_{n-2}I_{n}}|X,\mathbf{U}_{I_{n-2}})\\
 & +\sum_{k=1}^{n-2}\frac{1}{k}(H(U_{k}|\mathbf{Y}_{I_{k-1}I_{k}},\mathbf{U}_{I_{k-1}})-H(U_{k}|X,\mathbf{Y}_{I_{k-1}I_{n}},\mathbf{U}_{I_{k-1}})
\end{align*}
with additional information needed to encode repair information
\begin{align*}
R_{r} & >\frac{1}{n-1}\sum_{k=1}^{n-2}\left[H(Y_{kn}|\mathbf{U}_{I_{k}},\mathbf{Y}_{kI_{n-1}}\mathbf{Y}_{I_{k-1}I_{n}})\vphantom{-\frac{1}{n}H(\mathbf{Y}_{kI_{n}}|X,\mathbf{Y}_{k-1I_{n}},\mathbf{U}_{I_{k}})}\right.\\
 & \left.-\frac{1}{n}H(\mathbf{Y}_{kI_{n}}|X,\mathbf{Y}_{k-1I_{n}},\mathbf{U}_{I_{k}})\right]^{+}
\end{align*}
with $[x]^{+}=\max\{0,x\}$
\end{thm}
\begin{IEEEproof}
Here we will just outline how the coding changes compared to \cite{PuriRamchandranAl04PartI,PuriRamchandranAl05PartII},
and reading the proof therefore requires familiarity with those two
papers; the journal version will contain a formal proof. The formal
proof is in the Appendix. Consider first layer 1. We generate a random
codebook $\mathcal{C}_{u1}$ with $2^{lR_{u1}'}$ elements according
to the marginal distribution $p_{U_{1}}(u_{1})$. We also generate
$n$ independent random codebooks $\mathcal{C}_{1I_{n}}$ according
to the distribution $p_{Y_{11}}(y_{11})$ with $2^{lR_{1}'}$ codewords.
We first look for a codeword in $\mathcal{C}_{u1}$ that is jointly
typical with $x^{l}$ . Such a codeword can be found with high probability
if 
\begin{align*}
R_{u1}=R_{u1}' & >H(U_{1})-H(U_{1}|X)=I(X;U_{1})
\end{align*}
This codeword is stored in all the nodes. Next we look for an $n$-tuple
of codewords from $\mathcal{C}_{1I_{n}}$that are \emph{jointly} typical
with $x^{l}$ \emph{and} the chosen codeword from $\mathcal{C}_{u1}$.
These can be found with high probability if (compare \cite[(11)]{PuriRamchandranAl04PartI})
\begin{align}
nR_{1}' & >\sum_{i=1}^{n}H(Y_{1i})-H(Y_{11},\ldots,Y_{1n}|U_{1},X)\label{R1pproof.eq}
\end{align}
The codewords for each $Y_{1j}$ are binned into $2^{nR_{1}}$ bins.
At the time of decoding, the codeword for $U_{1}$ is available as
well as the bin number $i$ for $Y_{1j}$. The decoder looks for a
codeword in bin $i$ that is typical with $U_{1}$. If there is more
than one, the decoding results in error. If (compare \cite[(12)]{PuriRamchandranAl04PartI})
\begin{equation}
R_{1}'-R_{1}<H(Y_{11})-H(Y_{11}|U_{1})\label{R1pR1proof.eq}
\end{equation}
there is only one such codeword with high probability. Combining (\ref{R1pproof.eq})
and (\ref{R1pR1proof.eq}) we get
\[
R_{1}>H(Y_{11}|U_{1})-H(Y_{i1},\ldots,Y_{in}|U_{1},X)
\]

At layer $k<n-1$ we similarly generate a random codebook $\mathcal{C}_{uk}$
with $2^{lR_{uk}'}$ elements according to the marginal distribution
$p_{U_{k}}(u_{k})$ and $n$ independent random codebooks $\mathcal{C}_{kI_{n}}$
according to the distribution $p_{Y_{k1}}(y_{k1})$ with $2^{lR_{k}'}$
codewords. We first look for a codeword in $\mathcal{C}_{uk}$ that
is jointly typical with $x^{l}$ and all the codewords chosen in the
previous layers. This is possible with high probability if (compare
\cite[(12)]{PuriRamchandranAl05PartII}
\begin{align*}
R_{uk}' & >H(U_{k})-H(U_{k}|X,\mathbf{Y}_{I_{k-1}I_{n}},\mathbf{U}_{I_{k-1}})
\end{align*}
we then choose an $n$-tuple of codewords from $\mathcal{C}_{kI_{n}}$
that are jointly typical with all prior codewords and $x^{l}$ , which
is possible with high probability if (compare \cite[(12)]{PuriRamchandranAl05PartII}
\begin{align*}
nR_{k}' & >\sum_{i=1}^{n}H(Y_{ki})-H(\mathbf{Y}_{kI_{n}}|X,\mathbf{Y}_{k-1I_{n}},\mathbf{U}_{I_{k}})
\end{align*}
For $U_{k}$ we generate $n$ independent binning partitions each
with $2^{lR_{uk}}$ elements. The bin number in the $i$-th partition
is stored in the $i$-th node\footnote{As argued in \cite[Section III.E]{PuriRamchandranAl04PartI} this
kind of binning is equivalent to MDS erasure codes.}. The codewords for each of the $Y_{kj}$ are binned into $2^{lR_{k}}$
bins, and the bin number for $Y_{kj}$ stored in node $j$.

At the decoder, $k$ random nodes are used \textendash{} due to symmetry
we can assume these are the first $k$ nodes. We assume that the decoding
of the previous layers have been successful. So we know correctly
the codewords for $U_{1},\ldots,U_{k-1}$ , as well as $\mathbf{Y}_{I_{k-1}I_{k}}$,
and $k$ bin numbers for $U_{k}$ the bin number for each of $Y_{k1},\ldots,Y_{kk}$.

The decoder first looks for a \emph{common} codeword in the $k$ bins
for $U_{k}$ that is jointly typical with $(\mathbf{U}_{I_{k-1}},\mathbf{Y}_{I_{k-1}I_{k}})$.
With high probability there is only one such if (compare \cite[Section III.E]{PuriRamchandranAl04PartI})
\[
kR_{uk}>R_{uk}'+H(U_{k}|\mathbf{Y}_{I_{k-1}I_{k}},\mathbf{U}_{I_{k-1}})-H(U_{k})
\]
or
\begin{equation}
R_{uk}>\frac{1}{k}(H(U_{k}|\mathbf{Y}_{I_{k-1}I_{k}},\mathbf{U}_{I_{k-1}})-H(U_{k}|X,\mathbf{Y}_{I_{k-1}I_{n}},\mathbf{U}_{I_{k-1}}))\label{Rukproof.eq}
\end{equation}
It next looks in the $k$ bins for $Y_{k1},\ldots,Y_{kk}$ for codewords
that are jointly typical with $(\mathbf{U}_{I_{k}},\mathbf{Y}_{I_{k-1}I_{k}})$.
With high probability there is only one such is (compare \cite[(15)]{PuriRamchandranAl05PartII})
\[
k(R_{k}'-R_{k})<kH(Y_{k1})-H(\mathbf{Y}_{kI_{k}}|\mathbf{U}_{I_{k}},\mathbf{Y}_{I_{k-1}I_{k}})
\]
or
\[
R_{k}>\frac{1}{k}H(\mathbf{Y}_{kI_{k}}|\mathbf{U}_{I_{k}},\mathbf{Y}_{I_{k-1}I_{k}})-\frac{1}{n}H(\mathbf{Y}_{kI_{n}}|X,\mathbf{U}_{I_{k}},\mathbf{Y}_{I_{k-1}I_{n}})
\]
(as in \cite{PuriRamchandranAl04PartI} this can be repeated for any
collection of $k$ nodes).

At layer $n-1$ only a single codebook is generated, and this is binned
into $n$ independent partitions. Upon receipt, in analogy with (\ref{Rukproof.eq}),
this can be found uniquely with high probability if
\begin{align*}
R_{n-1} & >\frac{1}{n-1}H(Y_{n-1}|\mathbf{Y}_{I_{n-2}I_{n-1}},\mathbf{U}_{I_{n-2}})\\
 & -\frac{1}{n-1}H(Y_{n-1}|X,\mathbf{Y}_{I_{n-2}I_{n}},\mathbf{U}_{I_{n-2}})
\end{align*}

For repair, the joint $2^{nlR_{k}'}$ codewords in $\mathcal{C}_{k1}\times\cdots\times\mathcal{C}_{kn}$
at layer $k<n-1$ are binned into $2^{lR_{rk}}$ bins. The single
bin number of the $n$ chosen codewords is encoded with an $(n,n-1)$
MDS erasure code.

Now suppose node $n$ is lost, and needs to be recovered. The repair
works from the bottom up. So, suppose the bottom $k-1$ layers have
been recovered, that is, $\mathbf{Y}_{I_{k-1}I_{b}},\mathbf{U}_{I_{k-1}}$
are known without error. First $U_{k}$ is recovered, which can be
done since $n-1\geq k$ nodes are used. It can also decode the codewords
corresponding to $\mathbf{Y}_{kI_{n-1}}$. It restores the bin number
of the joint codeword from the erasure code. There are approximately
$2^{l(nR_{k}'-R_{rk})}$ codewords in the bin, but since it knows
the codewords of $\mathbf{Y}_{kI_{n-1}}$ there are only about $2^{l(R_{k}'-R_{rk})}$
valid ones. It searches in this bin for codewords jointly typical
with $\mathbf{Y}_{kI_{n-1}}\mathbf{Y}_{I_{k-1}I_{n}},\mathbf{U}_{I_{k}}$.
With high probability there is only one such if
\[
R_{k}'-R_{rk}<H(Y_{kn})-H(Y_{kn}|\mathbf{U}_{I_{k}},\mathbf{Y}_{kI_{n-1}}\mathbf{Y}_{I_{k-1}I_{n}})
\]
or
\begin{align}
R_{rk} & >H(Y_{kn}|\mathbf{U}_{I_{k}},\mathbf{Y}_{kI_{n-1}}\mathbf{Y}_{I_{k-1}I_{n}})\nonumber \\
 & -\frac{1}{n}H(\mathbf{Y}_{kI_{n}}|X,\mathbf{Y}_{k-1I_{n}},\mathbf{U}_{I_{k}})\label{Rrkbound.eq}
\end{align}
There is at least one codeword in the bin, namely the correct one.
Thus, if there is no error (more than one codeword), the repair is
exact, as required from the exact repairability condition in Section
\ref{Problem.sec}. 
\end{IEEEproof}

The above result can easily be adapted to the case of a repair node
that collaborates with the operational nodes. There are only two
differences
\begin{itemize}
\item The repair node can restore operation of the full $n$ node distortion
region. Therefore, the terminal single common codeword is not at layer
$n-1$ but at layer $n$. At the same time, the repair node now has
to store repair information for this last codeword. 
\item For distributed repair, distributions are chosen to minimize $R+R_{r}$.
For a repair node, distributions are chosen to minimize $R$, and
$R_{r}$ is then as given for those distributions.
\end{itemize}
With this in mind, we get

\begin{thm}[Collaborative repair node]
\label{CollaborativeRepair.thm}For any symmetric probability distribution
$p(\mathbf{y}_{I_{n-1},I_{n}},\mathbf{u}_{I_{n-1}},y_{n}|x)$ and
decoding functions $g_{J}$ the lower convex closure of $(R,D_{1},\ldots,D_{n})$
is achievable, where $E[d_{|J|}(X,g_{J}(\mathbf{Y}_{I_{|J|}J},\mathbf{U}_{I_{|J|}})]\leq D_{|J|},|J|\leq n$
and
\begin{align*}
\lefteqn{R>I(X;U_{1})+H(Y_{1n}|U_{1})+\sum_{k=2}^{n-1}\frac{1}{k}H(\mathbf{Y}_{kI_{k}}|\mathbf{Y}_{I_{k-1},I_{k}}\mathbf{U}_{I_{k}})}\\
 & +\frac{1}{n}H(Y_{n}|\mathbf{Y}_{I_{n-1}I_{n}},\mathbf{U}_{I_{n-1}})-\frac{1}{n}H(\mathbf{Y}_{I_{n-1}I_{n}}|X,\mathbf{U}_{I_{n-1}})\\
 & +\sum_{k=1}^{n-1}\frac{1}{k}(H(U_{k}|\mathbf{Y}_{I_{k-1}I_{k}},\mathbf{U}_{I_{k-1}})-H(U_{k}|X,\mathbf{Y}_{I_{k-1}I_{n}},\mathbf{U}_{I_{k-1}}))
\end{align*}
The additional information the repair node has to store is
\begin{align*}
R_{r} & >\sum_{k=1}^{n-1}\left[H(Y_{kn}|\mathbf{U}_{I_{k}},\mathbf{Y}_{kI_{n-1}}\mathbf{Y}_{I_{k-1}I_{n}})\vphantom{-\frac{1}{n}H(\mathbf{Y}_{kI_{n}}|X,\mathbf{Y}_{k-1I_{n}},\mathbf{U}_{I_{k}})}\right.\\
 & \left.-\frac{1}{n}H(\mathbf{Y}_{kI_{n}}|X,\mathbf{Y}_{k-1I_{n}},\mathbf{U}_{I_{k}})\right]^{+}\\
 & +\frac{1}{n}H(Y_{n}|\mathbf{Y}_{I_{n-1}I_{n}},\mathbf{U}_{I_{n-1}})-\frac{1}{n}H(\mathbf{Y}_{I_{n-1}I_{n}}|X,\mathbf{U}_{I_{n-1}})
\end{align*}
\end{thm}

\subsection{Example Gaussian case}

\begin{figure}[tbh]
\vspace{-0.1in}
\includegraphics[width=3.5in]{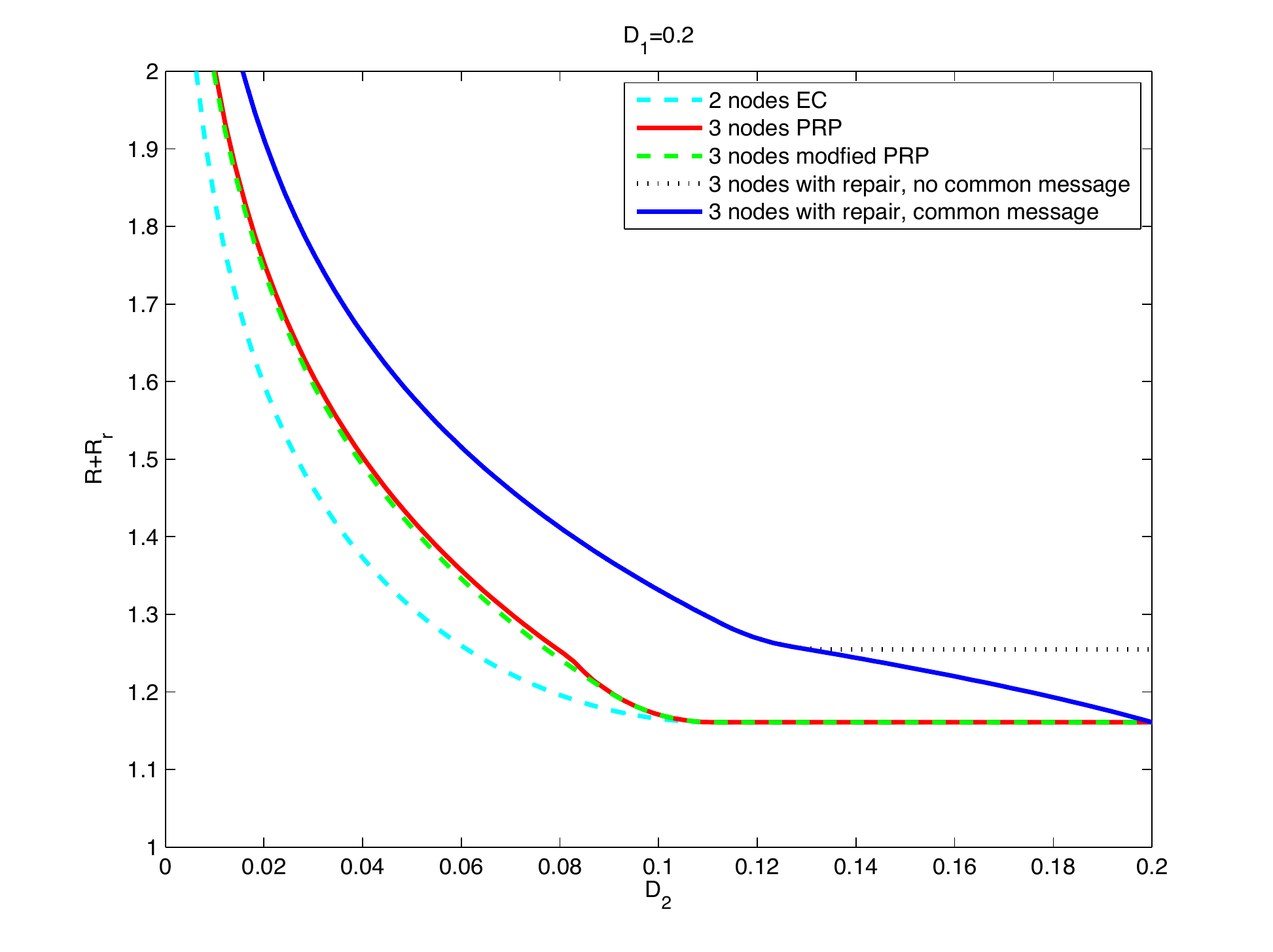}

\vspace{-0.2in}

\caption{\label{3node.fig}Plots of $R$ or $R+R_{r}$ for (a) two nodes according
to \cite{ElGamalCover82} (b) Three nodes with at most two used, without
repair according to PRP \cite{PuriRamchandranAl05PartII} or Theorem
\ref{DistributedRecov.thm} (modified PRP) and (c) Three nodes with
distributed repair without or without common message.}
\end{figure}
We consider a three node Gaussian case with mean-squared distortion
and with distributed repair. This is characterized by $(R+R_{r},D_{1},D_{2})$.
From Theorem \ref{DistributedRecov.thm} we get specifically (omitting
the $[]^{+}$operation)

\begin{align*}
R & \geq I(X;U_{1})+H(Y_{13}|U_{1})+\frac{1}{2}H(Y_{2}|Y_{13},Y_{12},U_{1})\\
 & -\frac{1}{3}H(Y_{13},Y_{12},Y_{11}|X,U_{1})-\frac{1}{2}H(Y_{2}|Y_{13},Y_{12},X,U_{1})\\
R_{r}\geq & \frac{1}{2}H(Y_{13}|Y_{12},Y_{11},U_{1})-\frac{1}{6}H(Y_{13},Y_{12},Y_{11}|X,U_{1})
\end{align*}
We put $U=X+Q_{U1},Y_{1i}=X+Q_{1i},Y_{2}=X+Q_{2}$ with $Q_{\ldots}$
zero-mean Gaussian, $E[Q_{u1}^{2}]=\sigma_{u1}^{2}$, $E[Q_{1i}^{2}]=\sigma_{q1}^{2}$,
$E[Q_{2}^{2}]=\sigma_{q2}^{2}$, $E[Q_{1i}Q_{1j}]=\rho_{1}\sigma_{q1}^{2}$,
and all other noise variables uncorrelated. Space does not allow us
to write down all details of the solution, but we will outline the
structure. The distortion and rates are calculated as in the proof
of Theorem \ref{2node.thm}. As in that proof there are three solution
regions: for small $D_{2}$ the nodes store resolution information,
but no common information. For medium $D_{2}$ the nodes store neither
resolution information nor common information; distortion $D_{2}$
is achieved solely through adjusting $\rho_{1}$. And for large $D_{2}$
the nodes store a common message, but no resolution information. In
all three regions it is possible to obtain closed form expressions
of rates, but they must be numerically optimized over $\rho_{1}$.

The distortions are still given by (\ref{D22node.eq}) and we still
have (\ref{sq1s.eq}). Then
\begin{align*}
R & =\frac{1}{2}\log\left(1+\frac{1}{\sigma_{u1}^{2}}\right)\\
 & +\frac{1}{2}\log\left(\frac{1+\sigma_{q1}^{2}-\frac{1}{1+\sigma_{u1}^{2}}}{\left((\rho_{1}-1)^{2}(2\rho_{1}+1)\right)^{1/3}\sigma_{q1}^{2}}\right)\\
 & +\frac{1}{4}\log\left(\frac{(\rho_{1}+1)\sigma_{q1}^{2}(\sigma_{q2}^{2}\sigma_{u1}^{2}+\sigma_{q2}^{2}+\sigma_{u1}^{2})+2\sigma_{q2}^{2}\sigma_{u1}^{2}}{((\rho_{1}+1)\sigma_{q1}^{2}(\sigma_{u1}^{2}+1)+2\sigma_{u1}^{2})\sigma_{q2}^{2}}\right)
\end{align*}
and
\begin{align*}
R_{r} & =\frac{1}{4}\log\left(\frac{(1-\rho_{1})\sigma_{q1}^{2}((2\rho_{1}+1)\sigma_{q1}^{2}(\sigma_{u1}^{2}+1)+3\sigma_{u1}^{2})}{\left((\rho_{1}+1)\sigma_{q1}^{2}(\sigma_{u1}^{2}+1)+2\sigma_{u1}^{2}\right)\left((\rho_{1}-1)^{2}(2\rho_{1}+1)\right)^{1/3}\sigma_{q1}^{2}}\right)
\end{align*}

Following the proof of Theorem \ref{2node.thm} we first consider
a solution with no resolution information, i.e, $\sigma_{q2}^{2}=\infty$.
Then we have
\[
D_{2}=\frac{D_{1}(\rho_{1}+1)\sigma_{u1}^{2}}{D_{1}(\rho_{1}-1)(\sigma_{u1}^{2}+1)+2\sigma_{u1}^{2}}
\]
And
\begin{align*}
R & =\frac{1}{2}\log\left(\frac{1}{D_{1}\sqrt[3]{(\rho_{1}-1)^{2}(2\rho_{1}+1)}}\right)\\
R_{r} & =\frac{1}{4}\log\left(\frac{(1-\rho_{1})(2D_{1}(\rho_{1}-1)(\sigma_{u1}^{2}+1)+3\sigma_{u1}^{2})}{\sqrt[3]{(\rho_{1}-1)^{2}(2\rho_{1}+1)}(D_{1}(r-1)(\sigma_{u1}^{2}+1)+2\sigma_{u1}^{2})}\right)
\end{align*}
For the solution without common information we get\footnote{This expression is only valid for such $\rho_{1}$ that $R_{r}>0$.}
\[
R_{t}=\frac{1}{2}\log\left(\frac{\sqrt{\frac{2D_{1}(\rho_{1}-1)+3}{\left(-2\rho_{1}^{2}+\rho_{1}+1\right)(D_{1}(\rho_{1}-1)+2)}}}{D_{1}}\right)
\]
This can be numerically minimized over $\rho_{1}\in(-\frac{1}{2},1)$.
But we also need the $D_{2}$ constraint satisfied, which requires
\[
\rho_{1}\leq\frac{D_{1}D_{2}+D_{1}-2D_{2}}{D_{1}(D_{2}-1)}
\]

Now, with common information, we aim to achieve exactly $D_{2}$,
so that
\[
\sigma_{u1}^{2}=\frac{D_{1}D_{2}(1-\rho_{1})}{2D_{2}-D_{1}(D_{2}(1-\rho_{1})+\rho_{1}+1)}
\]
This is valid for 
\[
\rho_{1}<\frac{2D_{2}-D_{1}D_{2}-D_{1}}{D_{1}(1-D_{2})}
\]
Then
\begin{align*}
R_{t} & =\frac{1}{2}\log\left(\frac{D_{2}\sqrt{\frac{D_{2}-2D_{1}(\rho_{1}+1)}{D_{1}(2\rho_{1}+1)}}}{D_{1}\rho_{1}+D_{1}-2D_{2}}\right)
\end{align*}
which must again by numerically optimized over $\rho_{1}$.

With resolution information we put $\sigma_{u1}^{2}=\infty$ and
\[
\sigma_{q2}^{2}=\frac{D_{1}D_{2}(\rho_{1}+1)}{D_{1}(D_{2}(-\rho_{1})+D_{2}+\rho_{1}+1)-2D_{2}}
\]
This is valid if
\[
\rho_{1}>\frac{2D_{2}-D_{1}D_{2}-D_{1}}{D_{1}(1-D_{2})}
\]
Then
\[
R_{t}=\frac{1}{2}\log\left(\frac{\sqrt{\frac{D_{1}(\rho_{1}+1)}{D_{2}(D_{1}(\rho_{1}-1)+2)}}}{D_{1}\sqrt[3]{(\rho_{1}-1)^{2}(2\rho_{1}+1)}}\right)
\]

Figure \ref{3node.fig} shows typical numerical results. First, from
the bottom, we have the curve for the optimum region for the two node
problem according to EC \cite{ElGamalCover82,ElGamalKimBook}. Notice
that this is achieved without any refinement information, using only
correlation between the base layer random variables; refinement information
is only required for $D_{1}>\frac{1}{2}$ and $D_{2}<2D_{1}-1$. Second,
we have the curves for the three node problem, but where we use at
most two nodes for reconstruction, either using \cite[Section V]{PuriRamchandranAl05PartII}
directly (ignoring the $D_{3}$ constraint), or using Theorem \ref{DistributedRecov.thm}
without repair. It can be noticed that using Proposition \ref{DistributedRecov.thm}
gives a slight improvement; this is \emph{not} due to the common message,
but due to the fact that PRP uses $n-1$ codewords in the last layer,
while the modified PRP uses only one. Finally, we have the curves
for repair. We see that a common message gives a clear improvement.

\section{Conclusion}

Our solutions show that it is sub-optimum for repair to just take
a standard multiple description code and add repair information. Rather,
the multiple description code has to be designed with repair in mind.
In this paper we do this by adding common messages. For the two node
case this was shown to be optimum. However, for the $n$ node case,
there might be better solutions.

\appendices{}

\section{Proof of Theorem \ref{DistributedRecov.thm}}

The proof is a modification of the proof of Theorem 2 in \cite{PuriRamchandranAl05PartII},
and we will therefore follow their steps closely. We let $T_{\epsilon}^{l}(X)$
denote the strongly $\epsilon$ typical set for $X$.

The coding scheme for repair uses MDS erasure codes in several places.
These can be put in the binning framework of PRP \cite{PuriRamchandranAl04PartI}.
However, it is easier to think of them as pure channel codes. We can
state this as follows: 
\begin{rem}
\label{MDS.thm}A message $M\in\{1,\ldots,2^{lR}\}$ is stored on
$n$ servers, of which at least $k$ is accessed for decoding. With
$R'>\frac{1}{k}R$ bits on each server, decoding is possible with
error $P(E)\to0$ as $l\to\infty$. 
\end{rem}

\subsection{Codebook generation}

The codebooks $\mathcal{C}_{I_{n-2}I_{n}}$ are generated and binned
exactly as in \cite{PuriRamchandranAl05PartII}. The difference from
\cite{PuriRamchandranAl05PartII} is that there is no $n$-th layer,
and that at layer $n-1$ there is only one codebook $\mathcal{C}_{n-1}$.
The codebook $\mathcal{C}_{n-1}$ of size $2^{lR'_{n-1}}$ is generated
like $\mathcal{C}_{n}$ in \cite{PuriRamchandranAl05PartII}, but
then stored on the nodes with an $(n,n-1)$ MDS code.

We also generate $n-2$ common codebooks $\mathcal{C}_{uI_{n-2}}$
by drawing $2^{lR'_{uk}}$ codewords $\left(\mathbf{u}_{k}^{(l)}(1),\ldots,\mathbf{u}_{k}^{(l)}(2^{lR'_{u_{k}}})\right)$
independently with replacement over the set $T_{\epsilon}^{l}(U_{k})$
according to a uniform distribution. The indices for $\mathcal{C}_{uk},k=2,\ldots,,n-2$
are next binned. Let $\xi_{uk}=2^{l(R'_{uk}-R_{uk}+\gamma_{uk})}$
for some $\gamma_{k}>0$ and make $2^{lR_{uk}}$ bins. For each bin,
select $\xi_{uk}$ numbers from the set $\{1,\ldots,2^{lR_{uk}'}\}$,
uniformly and with replacement. They are finally coded with an $(n,k)$
MDS erasure code. 

We finally generate $(n-1)$ repair codebooks through binning. First,
if
\begin{align}
0 & >H(Y_{kn}|\mathbf{U}_{I_{k}},\mathbf{Y}_{kI_{n-1}}\mathbf{Y}_{I_{k-1}I_{n}})\nonumber \\
 & -\frac{1}{n}H(\mathbf{Y}_{kI_{n}}|X,\mathbf{Y}_{k-1I_{n}},\mathbf{U}_{I_{k}})\label{NoRepair.eq}
\end{align}
it turns out, as will be seen later, that the lost codeword can be
recovered from the remaining ones with high probability. In that case
we set $R_{rk}=0$ and store no extra repair information. For consistency
we think of there being one bin at layer $k$ containing all codewords.
Otherwise, we let $\xi_{rk}=2^{l(nR'_{k}-R_{rk}+\gamma_{rk})}$ for
some $\gamma_{rk}>0$ and make $2^{lR_{rk}}$ bins. For each bin,
select $\xi_{rk}$ vectors from the set $\{1,\ldots,2^{lR_{k}'}\}^{n}$,
uniformly and with replacement. The bin indices are further coded
with an $(n,n-1)$ MDS erasure code.

\subsection{Encoding}

Given a source codeword $\mathbf{x}^{(l)}\in\mathcal{X}^{l}$ we find
codewords so that
\[
\left(\mathbf{x}^{(l)},\mathbf{u}_{I_{n-2}}^{(l)}(\mathbf{V}_{I_{n-2}}),\mathbf{y}_{I_{n-2}I_{n}}^{(l)}.*(\mathbf{Q}_{I_{n-2}I_{m}}),\mathbf{y}_{n-1}^{(l)}(Q_{n-1})\right)
\]
are jointly typical. The binning of $\mathbf{Q}_{I_{n-2}I_{m}}$ and
$Q_{n-1}$ are done exactly as in \cite{PuriRamchandranAl05PartII}
to obtain bin indices $B_{I_{n-2}I_{n}},B_{n-1}$. The bin index $B_{n-1}$
is further coded with the $(n,n-1)$ MDS code. For $V_{k}$ we find
the smallest bin index $B_{uk}$ that contains $V_{k}$ (if $V_{k}$
is in no bin, $B_{uk}=0$), and this is further coded with the $(n,k)$
MDS code.

For repair, for those $k\in I_{n-1}$ where repair information is
needed, we find the smallest bin index $W_{k}$ so that $\mathbf{Q}_{kI_{n}}$
is in the corresponding bin; if no bin contains $W_{k}$ we put $W_{k}=0$.
These are then coded with the $(n,n-1)$ MDS code.

\subsection{Decoding}

We assume node $1,2,\ldots,j$ are available. The bin indices $B_{uI_{j'}}$
are decoded from the MDS code, where $j'=\min\{j,n-2\}$. The decoding
now is similar to \cite{PuriRamchandranAl05PartII}, except that there
is also a common codeword. Consider decoding at layer $k\in\{2,\ldots,j'\}$.
First we find an index $V_{k}$ in bin $B_{uk}$ so that
\[
\left(\mathbf{y}_{I_{k-1}I_{j}}^{(l)},\mathbf{u}_{k}^{(l)}(V_{k}),\mathbf{u}_{I_{k-1}}^{(l)}\right)\in T_{\epsilon}^{l}\left(\mathbf{Y}_{I_{k-1}I_{j}},\mathbf{U}_{I_{k}}\right)
\]
Next, for any size $k$ subset $S\subset I_{j}$ the decoder looks
in bins $\mathbf{B}_{kS}$ for codewords $\mathbf{y}_{kS}^{(l)}$
so that
\[
\left(\mathbf{y}_{kS}^{(l)},\mathbf{y}_{I_{k-1}I_{j}}^{(l)},\mathbf{u}_{I_{k}}^{(l)}\right)\in T_{\epsilon}^{l}\left(\mathbf{Y}_{kI_{k}},\mathbf{Y}_{I_{k-1}I_{j}},\mathbf{U}_{I_{k}}\right)
\]

If $j=n-1$, $B_{n-1}$ is first recovered from the MDS code. Then
the above procedure is repeated (there is no $U_{n-1}$).

The reconstructions of $\hat{\mathbf{x}}^{(l)}$ is standard as in
\cite{PuriRamchandranAl05PartII}.

\subsection{Repair}

Without loss of generality and to simplify notation, we can assume
that node $n$ fails. The repair is done layer by layer. At layer
1, we copy $V_{1}$ from any node to the replacement node $n$. Next,
from the $(n-1)$ surviving nodes we decode the repair bin index $W_{1}$
from the MDS code; if there is no extra repair information we put
$W_{1}=1$. We know $\mathbf{Q}_{1I_{n-1}}$ from the surviving nodes.
In bin $W_{1}$ we look for an index $Q_{1n}$ so that the corresponding
codeword $(\mathbf{y}^{(l)}.*(\mathbf{Q}_{1I_{n}}),\mathbf{u}_{1}^{(l)}(V_{1}))\in T_{\epsilon}^{l}(\mathbf{Y}_{1I_{n}},U_{1})$
; if there is more than one, there is a repair error. We then store
the recovered $Q_{1n}$ in the replacement node $n$.

The following layers proceed in almost the same way. But now to recover
the common messeage $V_{k}$ we choose arbitrarily $k$ of the surviving
nodes and decode $V_{k}$ just as with usual operation. The decoded
$V_{k}$ is then encoded with the exact same MDS code and we store
the corresponding codeword on the replacement node $n$. We next find
an index $Q_{kn}$ in bin $W_{k}$ so that $(\mathbf{y}_{I_{k}I_{n}}^{(l)}.*(\mathbf{Q}_{I_{k}I_{n}}),\mathbf{u}_{I_{k}}^{(l)}(V_{I_{k}}))\in T_{\epsilon}^{l}(\mathbf{Y}_{I_{k}I_{n}},\mathbf{U}_{I_{k}})$
.

On the last layer we simply decode $Q_{n-1}$ from the surviving nodes
as usual, and then we re-encode with the same MDS code, and store
the recovered bin index on the new node $n$.

We notice that this repair is exact: the information on the restored
node is exactly the same as on the failed node, except if a repair
error happens.

\subsection{Analysis of Decoding Error}

We have some slightly modified error events compared to \cite{PuriRamchandranAl05PartII}
and some additional ones. We find it necessary to write these down
explicitly
\begin{enumerate}
\item $E_{0}$: $\mathbf{x}^{(l)}\notin T_{\epsilon}^{l}(X)$.
\item $E_{1}$: There exists no indices so that 
\begin{align*}
\left(\mathbf{x}^{(l)},\mathbf{u}_{I_{n-2}}^{(l)}(\mathbf{V}_{I_{n-2}}),\mathbf{y}_{I_{n-2}I_{n}}^{(l)}.*(\mathbf{Q}_{I_{n-2}I_{m}}),\right.\\
\left.\mathbf{y}_{n-1}^{(l)}(Q_{n-1})\right)\\
\in T_{\epsilon}^{l}(X,U_{I_{n-2}},Y_{I_{n-2}I_{n}},Y_{n-1})
\end{align*}
\item $E_{2}$: Not all the indices $(\mathbf{B}_{2I_{n}},\ldots,\mathbf{B}_{(n-2)I_{n}},B_{n-1})$
are greater than zero.
\item $E_{3}$: For some subset $S\subset I_{n}$ with $|S|=k\in\{2,\ldots,n-2\}$
there exists some other $\mathbf{Q}_{kS}'$ in bins $\mathbf{B}_{kS}$
so that\footnote{We use a slightly different notation for $E_{3}$ compared to \cite{PuriRamchandranAl05PartII},
which we think is clearer.}
\[
\left(\mathbf{y}_{kS}^{(l)}(\mathbf{Q}_{kS}'),\mathbf{y}_{I_{k-1}I_{j}}^{(l)},\mathbf{u}_{I_{k}}^{(l)}\right)\in T_{\epsilon}^{l}\left(Y_{kI_{k}},Y_{I_{k-1}I_{k}},U_{I_{k}}\right)
\]
\item $E_{4}$: not all the indices $B_{uk}$ are greater than zero.
\item $E_{5}$: For some $2\leq k\leq n-2$ there exist another index $V_{k}'\neq V_{k}$
in bin $B_{uk}$ so that
\begin{equation}
\left(\mathbf{y}_{I_{k-1}I_{j}}^{(l)},\mathbf{u}_{k}^{(l)}(V_{k}'),\mathbf{u}_{I_{k-1}}^{(l)}\right)\in T_{\epsilon}^{l}\left(Y_{I_{k-1}I_{k}},U_{I_{k}}\right)\label{UkJT.eq}
\end{equation}
\item $E_{6}$: There is a decoding error in the $(n,k)$ MDS erasure code
for $B_{uk}$.
\item $E_{7}$: There is a decoding error in the $(n,n-1)$ MDS erasure
code for $B_{n-1}$.
\end{enumerate}
First by Remark \ref{MDS.thm}, $P(E_{6}),P(E_{7})\to0$ as long as
the rates before the MDS is scaled appropriately.

As in \cite{PuriRamchandranAl05PartII} we have $P(E_{0})\to0$ as
$l\to\infty$. For $E_{1}$ as in \cite{PuriRamchandranAl05PartII}
we define $E_{1i}$ as an encoding error on layer $i$ given that
the previous layers have been encoded correctly and \emph{in addition},
here, that $\mathbf{u}_{i}^{(l)}$ has been encoded correctly. Then
as in \cite{PuriRamchandranAl05PartII} we get that $P(E_{1i})\to0$
if
\begin{align}
nR_{1}' & >nH(Y_{11})-H(\mathbf{Y}_{1I_{n}}|X,U_{1})\nonumber \\
nR_{i}' & >nH(Y_{i1})-H(\mathbf{Y}_{iI_{n}}|X,\mathbf{Y}_{I_{i-1}I_{n}},\mathbf{U}_{I_{i-1}})\nonumber \\
nR_{n-1}' & >I(Y_{n-1};X,\mathbf{Y}_{I_{n-2}I_{n}}\mathbf{U}_{I_{n-2}})\label{Rkpproof.eq}
\end{align}
with the difference being the addition of the $U_{*}$ variables.
Similarly, we can define $E_{1i}^{u}$ as an encoding error of $\mathbf{u}_{i}^{(l)}$
given that the previous layers have been encoded correctly, and we
similarly have that $P(E_{1i}^{u})\to0$ if 
\begin{align}
R_{u1}' & >H(U_{1})-H(U_{1}|X)\nonumber \\
R_{ui}' & >H(U_{i})-H(U_{i}|X,\mathbf{Y}_{I_{i-1}I_{n}},\mathbf{U}_{I_{i-1}})\label{Rukpproof.eq}
\end{align}
The proof that $P(E_{2})\to0$ is unchanged from \cite{PuriRamchandranAl05PartII},
and the proof that $P(E_{4})\to0$ is similar.

The proof that $P(E_{3})\to0$ is similar to \cite{PuriRamchandranAl05PartII},
except that at the time of decoding at layer $k$ the decoder has
access to $\mathbf{u}_{I_{k}}^{(l)}$. The relevant probability of
decoding error at layer $k$ therefore is $P(E_{3k}|\mathbf{E}_{3I_{k-1}}^{c},E_{2}^{c},E_{4}^{c},\mathbf{E}_{5I_{k}}^{c},E_{6}^{c},E_{7}^{c})$,
and since we search for codewords in $T_{\epsilon}^{l}\left(Y_{kI_{k}},Y_{I_{k-1}I_{j}},U_{I_{k}}\right)$,
the condition for this error probability converging to zero is 

\begin{equation}
R_{k}>R_{k}'-H(Y_{k1})+\frac{1}{k}H(\mathbf{Y}_{kI_{k}}|\mathbf{U}_{I_{k}},\mathbf{Y}_{I_{k-1}I_{k}})\label{Rkproof.eq}
\end{equation}
instead of \cite[(A17)]{PuriRamchandranAl05PartII}.

To prove that $P(E_{5})\to0$ we let $E_{5k}$ be the decoding error
on layer $k$, and then bound $P_{5k}=P(E_{5k}|\mathbf{E}_{3I_{k-1}}^{c},E_{2}^{c},E_{4}^{c},\mathbf{E}_{5I_{k-1}}^{c},E_{6}^{c},E_{7}^{c})$.
If we pick a random codeword $\mathbf{u}_{k}^{(l)}\in T_{\epsilon}^{l}(U_{k})$
the probability that this is jointly typical, i.e., the event (\ref{UkJT.eq}),
is
\[
P\leq2^{-l(I(U_{k};Y_{I_{k-1}}U_{I_{k-1}})-\delta(\epsilon))}
\]
There are $\xi_{uk}=2^{l(R'_{uk}-R_{uk}+\gamma_{uk})}$ elements in
each bin, and therefore,
\[
P_{5k}\leq\xi_{uk}P
\]
if we let $\gamma_{uk}>\delta(\epsilon)$ we have $P_{5k}\to0$ if
\[
R'_{uk}-R_{uk}<I(U_{k};Y_{I_{k-1}}U_{I_{k-1}})
\]
Together with (\ref{Rukpproof.eq}) this gives (\ref{Rukproof.eq}).

\subsection{Analysis of Repair error}

If $E_{4-7}$, from above happen, there is also a repair error. Notice
that at time of repair, we have access to $n-1$ nodes, and we can
therefore use decoding for $n-1$ nodes, and in that case we have
proven that $\sum_{i=4}^{7}P(E_{i})\to0$ as $l\to\infty$. We have
the following additional repair error events
\begin{enumerate}
\item $E_{r1}$: Some $W_{k}=0$ for $k\in I_{n-2}$.
\item $E_{r2}$: For $k\in I_{n-2}$ there exists another bin index $Q_{kn}'$
in bin $W_{k}$ so that 
\begin{align*}
(\mathbf{y}_{kI_{n}}^{(l)}(\mathbf{Q}_{kI_{n}}'),\mathbf{y}_{I_{k-1}I_{n}}^{(l)}.*(\mathbf{Q}_{I_{k-1}I_{n}}),\mathbf{u}_{I_{k}}^{(l)}(V_{I_{k}}))\\
\in T_{\epsilon}^{l}(Y_{I_{k}I_{n}},U_{I_{k}})
\end{align*}
\item $E_{r3}$: For $k\in I_{n-2}$ there is a decoding error in the $(n,n-1)$
MDS erasure code for $W_{k}$.
\end{enumerate}

\subsubsection{Bounding $E_{r1}$}

In total, for all bins, we pick $N=2^{lR_{rk}}\xi_{rk}=2^{l(nR_{k}'+\gamma_{rk})}$
elements with replacement from a set of size $2^{nlR_{rk}'}$. The
probability that a particular element was never picked then is $P(E_{r1})=\left(1-2^{-nlR_{rk}'}\right)^{N}$
and
\begin{align*}
\log P(E_{r1}) & =N\log\left(1-2^{-nlR_{rk}'}\right)\leq-N2^{-nlR_{rk}'}\\
 & =-2^{l\gamma_{rk}}\to-\infty\quad\text{as }l\to\infty
\end{align*}

\subsubsection{Bounding $E_{r2}$}

First we will argue that if (\ref{NoRepair.eq}) is satisfied, we
can predict $\mathbf{y}_{kn}^{(l)}$ with probability approaching
one. We can state this as follows: if we pick a random $\mathbf{y}_{kn}^{(l)}\in T_{\epsilon}^{l}(Y_{kn})$,
what is the probability $P$ that 
\begin{align}
(\mathbf{y}_{kn}^{(l)},\mathbf{y}_{kI_{n-1}}^{(l)}(\mathbf{Q}_{kI_{n-1}}),\mathbf{y}_{I_{k-1}I_{n}}^{(l)}.*(\mathbf{Q}_{I_{k-1}I_{n}}),\mathbf{u}_{I_{k}}^{(l)}(V_{I_{k}}))\nonumber \\
\in T_{\epsilon}^{l}(Y_{I_{k}I_{n}},U_{I_{k}})\label{TypicalBin.eq}
\end{align}
This is actually a standard channel coding problem, so we get
\begin{equation}
P\leq2^{-l(I(Y_{kn};Y_{kI_{n-1}},Y_{iI_{n-1}},U_{I_{k}})-\delta(\epsilon))}\label{Ptypical.eq}
\end{equation}
Since the codebook $\mathcal{C}_{uk}$ has $2^{lR_{k}'}$ elements,
we then have 
\[
P(E_{r2k})\leq2^{lR_{k}'}P
\]
Thus, $P(E_{r2k})\to0$ as $l\to\infty$ if 
\[
R_{k}'<H(Y_{kn})-H(Y_{kn}|Y_{kI_{n-1}},Y_{iI_{n-1}},U_{I_{k}})-\delta(\epsilon)
\]
Now in consideration of (\ref{Rkproof.eq}) there is no gain from
making $R_{k}'$ larger than needed. Thus $R_{k}'$ is chosen arbitrarily
close to the limit given by (\ref{Rkpproof.eq}), and we therefore
have $P(E_{r2k})\to0$ if
\begin{align*}
 & H(Y_{k1})-\frac{1}{n}H(\mathbf{Y}_{kI_{n}}|X,\mathbf{Y}_{I_{k-1}I_{n}},\mathbf{U}_{I_{k-1}})\\
 & <H(Y_{kn})-H(Y_{kn}|Y_{kI_{n-1}},Y_{iI_{n-1}},U_{I_{k}})-\delta(\epsilon)
\end{align*}
which is (\ref{NoRepair.eq}).

Now turn to the case when (\ref{NoRepair.eq}) is not satisfied. We
look for vectors $(Q_{k1}',Q_{k2}',\ldots,Q_{kn}')\in\{1,\ldots,2^{lR_{k}'}\}^{n}$
that
\begin{enumerate}
\item Are in the bin indicated by $W_{k}$.
\item Has $Q_{ki}'=Q_{ki}$, $i\leq n-1$.
\item Are jointly typical, i.e., satisfy (\ref{TypicalBin.eq}).
\end{enumerate}
For condition 3, (\ref{Ptypical.eq}) is still valid. Each bin contains
$\xi_{rk}=2^{l(nR'_{k}-R_{rk}+\gamma_{rk})}$ vectors. Each of these
has probability $P_{2}=2^{-l(n-1)R_{lk}}$ of satisfying conditions
2. Therefore
\[
P(E_{r2k})\leq\xi_{rk}P_{2}P=2^{l(R'_{k}-R_{rk}+\gamma_{rk})}P
\]
if we choose $\gamma_{rk}>\delta(\epsilon)$ we have $P(E_{r2k})\to0$
as $l\to\infty$ if 
\[
R_{k}'-R_{rk}<H(Y_{kn})-H(Y_{kn}|Y_{kI_{n-1}},Y_{iI_{n-1}},U_{I_{k}})
\]
Which together with (\ref{Rkpproof.eq}) and the argument above leads
to (\ref{Rrkbound.eq}).

\bibliographystyle{IEEEtran}
\bibliography{BigData,Coop03,Coop06}

\end{document}